\documentclass[aps,prl,amsmath,amssymb,floatfix,twocolumn,groupedaddress,nofootinbib,10pt]{revtex4-1}
 
\usepackage{times}
\usepackage{bbold}
\usepackage{bm}
\usepackage{graphicx} 
\usepackage{hyperref}
\usepackage{textcomp}
\usepackage{amssymb} 
\usepackage{wasysym}
\usepackage{epstopdf}
\usepackage{pdfpages} 

\usepackage{multibib} 
\newcites{supp}{Supplemental References}

\makeatletter
\AtBeginDocument{\let\LS@rot\@undefined}
\makeatother
 
\begin{document}

\title{Majorana Kramers pairs in higher-order topological insulators}
 
\author{Chen-Hsuan Hsu$^{1}$}
\author{Peter Stano$^{1,2,3}$}
\author{Jelena Klinovaja$^{1,4}$}
\author{Daniel Loss$^{1,4}$}

\affiliation{$^{1}$RIKEN Center for Emergent Matter Science (CEMS), Wako, Saitama 351-0198, Japan}
\affiliation{$^{2}$Department of Applied Physics, School of Engineering, University of Tokyo, 7-3-1 Hongo, Bunkyo-ku, Tokyo 113-8656, Japan}
\affiliation{$^{3}$Institute of Physics, Slovak Academy of Sciences, 845 11 Bratislava, Slovakia}
\affiliation{$^{4}$Department of Physics, University of Basel, Klingelbergstrasse 82, CH-4056 Basel, Switzerland}

\date{\today}
                                                     
\begin{abstract}
We propose a tune-free scheme to realize Kramers pairs of Majorana bound states in recently discovered higher-order topological insulators (HOTIs). We show that, by bringing two hinges of a HOTI into the proximity of an $s$-wave superconductor, the competition between local and crossed Andreev pairing leads to the formation of Majorana Kramers pairs, when the latter pairing dominates over the former. We demonstrate that such a topological superconductivity is stabilized by moderate electron-electron interactions. The proposed setup avoids the application of a magnetic field or local voltage gates, and requires weaker interactions compared with nonhelical nanowires. 

\end{abstract}

\maketitle

Majorana bound states (MBSs), with topological quantum computation prospects, have gained much attention recently~\cite{Kitaev:2001,Kitaev:2003,Nayak:2008,Tanaka:2009,Sato:2009,Hasan:2010,Qi:2011,Alicea:2011,Cook:2011,Klinovaja:2012b,Chevallier:2012,Dominguez:2012,Klinovaja:2012c,Niu:2012,Prada:2012,Terhal:2012,Beenakker:2013,Bjornson:2013,Nakosai:2013,Thakurathi:2013,Dumitrescu:2014,Maier:2014,Weithofer:2014,Deacon:2017,Huang:2017,Izumida:2017,Manousakis:2017,Ptok:2017,Sato:2017,Reeg:2018,Trif:2018}. 
However, the prototypical realizations based on proximity-induced superconductivity and either semiconducting nanowires with strong spin-orbit interactions~\cite{Lutchyn:2010,Oreg:2010,Mourik:2012,Das:2012,Deng:2012,Rokhinson:2012,Finck:2013,Churchill:2013,Rainis:2013,Takei:2013,Albrecht:2016,Gul:2018,Zhang:2018}, or topological insulators (TIs)~\cite{Fu:2008,Fu:2009}
require an external magnetic field, detrimental to superconductivity and MBSs themselves. 
It might also be noted that buried Dirac points are common in two-/three-dimensional TIs (2DTIs/3DTIs)~\cite{Skolasinski:2017,Zhang:2009,Hsieh:2009,Brune:2011,Li:2018}, impeding the realization of MBSs in TI-superconductor heterostructures using magnetic fields~\cite{Skolasinski:2017}.

Platforms without magnetic fields are therefore searched for~\cite{Keselman:2013,Haim:2014,Schrade:2015,Wong:2012,Zhang:2013,Hoffman:2016b,Yan:2018}, examples including helical spin textures~\cite{Klinovaja:2013,Braunecker:2013,Vazifeh:2013,Nadj-Perge:2013,Nadj-Perge:2014,Pientka:2014,Hsu:2015,Pawlak:2016} and crossed Andreev pairings in double nanowires or 2DTI edge channels~\cite{Gaidamauskas:2014,Klinovaja:2014,Klinovaja:2014b,Klinovaja:2015,Ebisu:2016,Schrade:2017,Thakurathi:2018}. 
In the former, the spin texture arises through indirect coupling mediated by itinerant carriers. The superconducting gap reduces it, leading to a tradeoff between the operation temperature (set by the indirect coupling) and the MBS localization length (set by the superconducting gap). 
On the other hand, the latter setup requires fine-tuned chemical potentials in two isolated one-dimensional channels. These difficulties motivate us to seek a new scheme to avoid fine-tuning.

Here we propose such a scheme, exploiting the recently discovered higher-order topological insulators (HOTIs)~\cite{Benalcazar:2017,Benalcazar:2017b,Schindler:2017,Langbehn:2017,Song:2017,Ezawa:2018,Schindler:2018,Khalaf:2018,Beenakker:2018}. 
Specifically, we focus on 3D helical second-order TIs. 
In contrast to the gapless surface states in their first-order counterparts~\cite{Fu:2007,Moore:2007,Hsieh:2008,Roy:2009,Hasan:2010,Qi:2011}, these HOTIs host helical hinge states, in which opposite spins move in opposite directions, akin to the spin-momentum locked edge channels in 
2DTIs~\cite{Kane:2005b,Kane:2005,Bernevig:2006b,Bernevig:2006,Murakami:2006,Liu:2008,Wada:2011}. 
Important for us, these hinges form one-dimensional channels of identical chemical potentials. 
There is compelling experimental evidence for the topological hinge states in Bi(111) nanowires and bilayers~\cite{Drozdov:2014,Murani:2017,Schindler:2018}.

Our scheme exploits $s$-wave superconductivity-proximitized helical hinges of a HOTI. Two types of pairings arise, a local (standard) and a nonlocal (crossed Andreev) one. We first demonstrate that 
Majorana Kramers pairs (MKPs) emerge when the crossed Andreev pairing dominates. This regime, however, does not arise in a noninteracting system. Nevertheless, we show that rather weak electron-electron interactions are sufficient to push the system into a regime where the crossed Andreev pairing dominates. We therefore predict that MKPs typically appear at the ends of a HOTI nanowire. 

To elucidate an essential feature, consider that 
two parallel hinges of a helical HOTI are in contact with an $s$-wave superconductor. Cooper pairs can tunnel into the hinges through two processes. The local (nonlocal) pairing process corresponds to the two partners of a Cooper pair tunneling into the same (different) hinge(s). We denote the configuration as {\it parahelical} ({\it orthohelical}), when the helicities of the two hinges are {\it the same} ({\it opposite}). For example, in the parahelical setup spin-down electrons in the two hinges propagate in the same direction, whereas in the orthohelical setup they move in opposite directions. The momentum conservation imposes selection rules: the chemical potentials of the two hinges have to be the same (opposite) for the parahelical (orthohelical) setup~\cite{Klinovaja:2014} to allow for a crossed Andreev pairing. Since the conducting hinges of a HOTI are all connected, their chemical potentials are identical {\it without applying local voltage gates}, a substantial advantage.

\begin{figure}[t]
\includegraphics[width=0.45\linewidth]{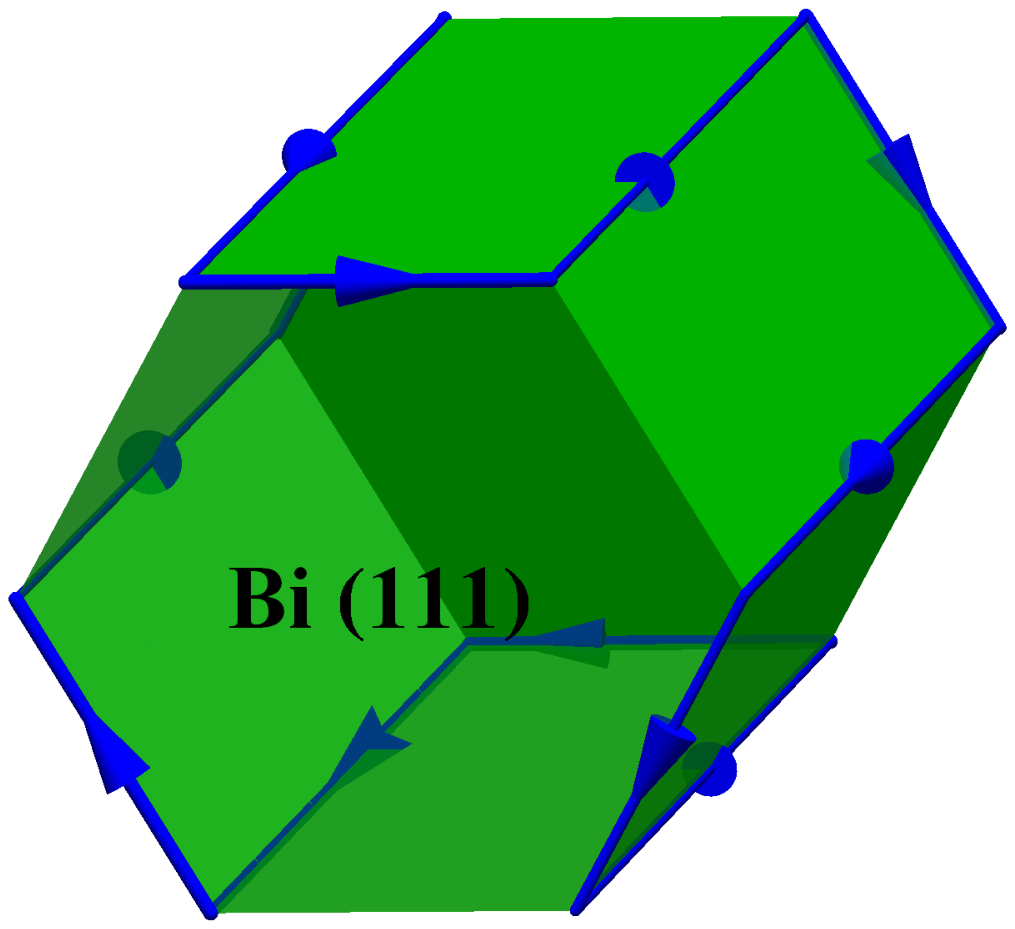}
\includegraphics[width=0.52\linewidth]{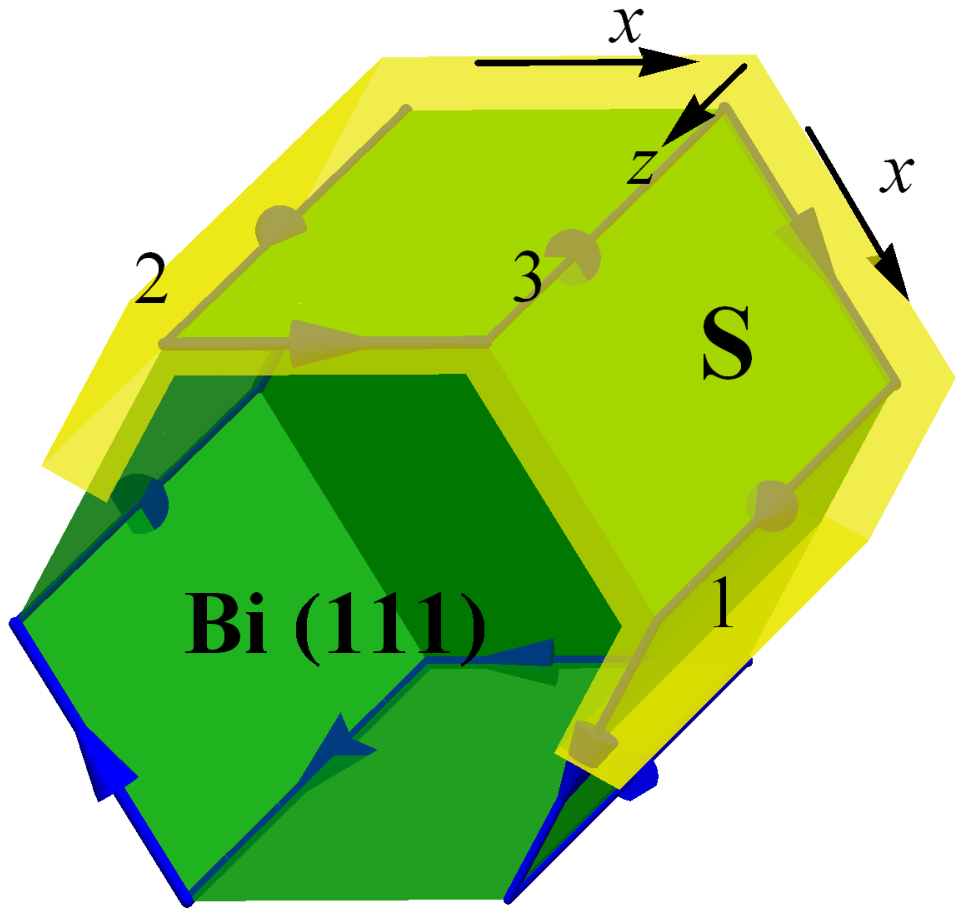}
 \caption{Left: In a Bi(111) nanowire (green), the gapless states (blue arrows) propagate along the hinges. The spin-up (-down) hinge states move against (along) the directions of the arrows. The helicities of any two parallel hinge states [along $z \equiv (111)$ axis] on a single lateral facet are opposite. 
Right: In the proposed setup, a superconducting layer (yellow) covers three parallel hinges (labeled by 1, 2, and 3). The (1,2) pair carries the same helicity, allowing for the crossed Andreev pairing. For other pairs, [(1,3) and (2,3)], such pairing is suppressed. The $x$ axis of the local coordinate is defined along the perimeter of the hexagonal cross section, and the $y$ axis (not shown) is normal to the lateral facets.}
\label{Fig:Setup}
\end{figure}

{\it Setup.} 
As a concrete example, we consider the recently discovered HOTI material, a bismuth crystal grown along the (111) axis,\footnote{Even though bismuth is a bulk semimetal, the topologically trivial bulk states can be gapped by, e.g., superconductivity, disorder or finite-size effects~\cite{Murakami:2006,Wada:2011,Murani:2017,Schindler:2018,Beenakker:2018}. 
Further, while we take Bi(111) nanowires as an example, our setup can be implemented also in other recently predicted helical HOTI materials, such as SnTe, Bi$_2$TeI, BiSe, and BiTe~\cite{Schindler:2017}.}  
which hosts helical hinge states~\cite{Schindler:2018}, as drawn in Fig.~\ref{Fig:Setup}.
Since the helicities of any two parallel hinges on the same lateral facet are opposite, the orthohelical setup is realized when a superconductor covers one lateral facet with two parallel hinges. In this case, crossed Andreev pairing is not feasible.

\begin{figure}[t]
 \includegraphics[width=0.95\linewidth]{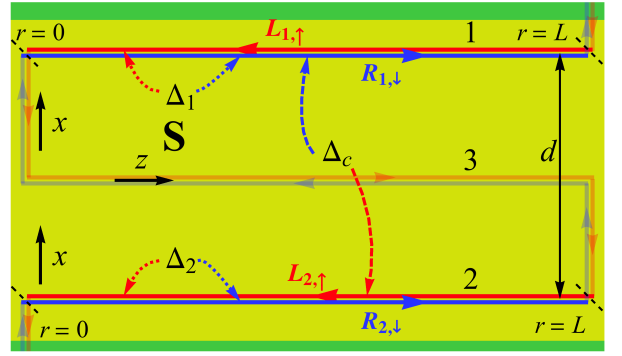}
 \caption{Schematics of the parahelical setup in the $xz$ plane of the local coordinate (view from the $+y$ direction) with the hinge coordinate $r$. A superconductor (yellow) covers three long hinges (along the $z$ direction), and several short hinges (along the $x$ direction). In hinges 1 and 2, which are at distance $d$, the spin-up states propagate toward the $-r$ direction (red solid arrows) while the spin-down states toward $+r$ (blue solid arrows). Hinge 3 is decoupled from the others. The local ($\Delta_{1,2}$) and crossed Andreev ($\Delta_{\rm c}$) pairing processes are indicated by the dotted and dashed arrows, respectively. 
Since the short segments along the $x$ direction are not aligned in the laboratory frame, $\Delta_{\rm c}$ is suppressed if $r \notin [0,L]$, while $\Delta_{1,2}$ remains constant for any $r$, including in the short segments.
 As a result, the boundaries (black dashed lines) are created at $r=0$ and $r=L$ (the ends of the nanowire), which are assumed to be far apart
on the scale of the Majorana localization length.
For clarity, only one crossed Andreev pairing process, $\Delta_{\rm c} R_{1,\downarrow}^\dagger L^\dagger_{2,\uparrow}$, is depicted.
}
\label{Fig:Parahelical}
\end{figure}

However, when the superconductor extends over two lateral facets (see the right panel of Fig.~\ref{Fig:Setup}), it is in contact with three parallel hinges along $z \equiv (111)$ axis. Two of them (labeled by 1 and 2) carry the same helicity while the third one (labeled by 3) the opposite. The momentum selection rules, together with uniform chemical potential (assumed to be in the bulk gap and not very close to the Dirac point) allow a crossed Andreev pairing between the hinges 1 and 2 and forbid it between the orthohelical hinges [(1,3) and (2,3) pairs], see Fig.~\ref{Fig:Parahelical}. As a result, the hinge 3 decouples from the remaining two and the parahelical setup is realized in the hinges 1 and 2 ~\cite{SM}.

{\it Model.} 
From now on we restrict ourselves to the two coupled hinges.\footnote{We consider a clean system with identical Fermi wave numbers for the two identical hinges. However, our results are robust against weak disorder~\cite{SM}.}
We model them using the fields
\begin{align}
\psi_n (r) =  R_{n, \downarrow} (r) e^{i k_F r} + L_{n, \uparrow}(r)  e^{-i k_F r},
\end{align} 
with the coordinate $r$ along the hinge, the hinge index $n \in \{1,2\}$, the Fermi wave number $k_F$, and the slowly varying right- and left-moving fields $R_{n, \downarrow}$ and $L_{n, \uparrow}$, respectively. Whenever possible, we suppress the coordinate $r$ and the spin index, the latter fixed by the spin-momentum locking. In a noninteracting system, the effective Hamiltonian reads 
$H = H_0 + H_{\rm intra} + H_{\rm c}$. The kinetic energy term is 
\begin{align}
H_0 = -i \hbar v_F \sum_{n=1,2} \int dr \; \Big( R_n^\dagger \partial_r R_{n} - L^\dagger_{n} \partial_r L_{n} \Big),
\label{Eq:H0}
\end{align}
with the Fermi velocity $v_F$. The local pairing term is
\begin{align}
H_{\rm intra} = \sum_{n=1,2} \int dr \; \Big[ \frac{\Delta_{n}}{2} (R_n^\dagger L^\dagger_{n} - L^\dagger_{n} R_{n}^\dagger) + {\rm H.c.} \Big],
\label{Eq:H_intra}
\end{align}
with the pairing gap $\Delta_{n}$ in the hinge $n$. 
Finally, the crossed Andreev pairing term is
\begin{align}
H_{\rm c} =  \int dr \;  \left[  \frac{\Delta_{\rm c}}{2}(R_1^\dagger L^\dagger_{2} - L^\dagger_{2} R_{1}^\dagger) + (1 \leftrightarrow 2) \right] + {\rm H.c.} , 
\label{Eq:H_cross}
\end{align}
with the pairing gap $\Delta_{\rm c}$. 
For simplicity, we take a spatially uniform real local pairing gap $\Delta_{n}>0$. On the other hand, the crossed Andreev pairing gap $\Delta_{\rm c}$ changes its (real) value from finite ($r \in [0,L]$) to zero ($r \notin [0,L]$),  
creating two boundaries at $r=0$ and $r=L$, as indicated in Fig.~\ref{Fig:Parahelical}. Assuming the hinge length $L$ being sufficiently long, we focus only on the boundary at $r=0$ and demonstrate the existence of a MKP localized there. 

{\it Majorana Kramers pairs.} 
We first identify the criterion for MKPs in a noninteracting system. 
The single-particle Hamiltonian, see Eqs.~(\ref{Eq:H0})--(\ref{Eq:H_cross}), can be written in the basis $\Psi = (R_1, L_1, R_2, L_2, R_1^\dagger, L_1^\dagger, R_2^\dagger, L_2^\dagger)^{\rm T}$ as $H_{\rm  } = \frac{1}{2} \int dr \; \Psi^{\dagger} (r) \mathcal{H}_{\rm  } (r) \Psi (r) $, with the Hamiltonian density
\begin{align}
\mathcal{H}_{\rm  }(r) =& -i \hbar v_F \eta^0 \tau^0 \sigma^z \partial_r  -\Delta_{+} \eta^y \tau^0  \sigma^y \nonumber \\
 &-\Delta_{-}  \eta^y \tau^z \sigma^y  - \Delta_{\rm c}  \eta^y \tau^x \sigma^y,
\label{Eq:H_para}
\end{align}
with $\Delta_{\pm} = (\Delta_1 \pm \Delta_2)/2$. In the above, the matrices $\eta^{\mu}$, $\tau^{\mu}$, and $\sigma^{\mu}$ act on the particle-hole, hinge, and spin space, respectively. They are given by the Pauli (identity) matrix for $\mu = x,y,z$ ($\mu =0$). 
The bulk spectrum is two-fold degenerate due to the time-reversal symmetry (TRS), with a gap denoted as $\Delta_{\rm b}$.
The reversal of the sign of $\Delta_{\rm b}$, which can be shown to coincide with the sign of $(\Delta_1 \Delta_2 - \Delta_{\rm c}^2)$, indicates the band inversion and suggests the presence of zero-energy MBSs at a boundary.
 
By directly solving the Bogoliubov--de Gennes equation of Eq.~\eqref{Eq:H_para} at zero energy~\cite{Klinovaja:2012a,Hsu:2015}, one can show that such bound states are indeed present ~\cite{SM}.
With this procedure, we find that a MKP at $r=0$ (and another pair at $r=L$) emerges if and only if
\begin{align}
\Delta_{\rm c}^2 > \Delta_{1} \Delta_{2}.
\label{Eq:criterion}
\end{align} 
Because of its topological origin, the MKP appears and disappears wherever $\Delta_{\rm b}$ reverses its sign even in setups with different model details, for example, a less abrupt change of $\Delta_{\rm c}$. 
Similarly, additional second-order (co-)tunneling processes~\cite{Klinovaja:2014,Reeg:2017,Schrade:2017} not included here do not
affect the topological criterion as long as the local pairing gaps $\Delta_n$ are of similar strengths~\cite{SM}.
We therefore conclude that the criterion for MKPs is the crossed Andreev pairing to be dominant over the local pairing, as described by Eq.~\eqref{Eq:criterion}. In noninteracting systems, however, Eq.~\eqref{Eq:criterion} cannot be fulfilled~\cite{Reeg:2017}. Including electron-electron interactions is thus essential for our scheme. Below we demonstrate that even moderate interactions can drive the system into the topological superconducting phase hosting MKPs. 

{\it Interacting system.}
To begin, we note that since our setup respects TRS, the elastic backscattering is precluded in the helical channels (unless the 
TRS is broken, for example, by nuclear spins~\cite{Hsu:2017,Hsu:2018}). We therefore include only the forward scattering processes into the interaction $H_{\rm int}$ and bosonize the total hinge Hamiltonian $\mathit{H}_{\textrm{el}} = \mathit{H}_{0} + H_{\rm int}$. This procedure leads to two copies of the helical Tomonaga-Luttinger liquid,
\begin{align}
\mathit{H}_{\textrm{el}} = \sum_{n=1,2} \int \frac{\hbar dr}{2\pi} \, \left\{ u_n K_n \left[ \partial_{r} \theta_n(r) \right]^2 + \frac{u_n}{K_n} \left[ \partial_{r} \phi_n(r) \right]^2 \right\},
\label{Eq:H_el}
\end{align}
with the interaction parameter $K_{n}$ for the hinge $n$ and the modified velocity $u_{n} = v_F/K_n$.
Using standard bosonic fields $\theta_n$ and $\phi_n$ for helical channels~\cite{Hsu:2017,Hsu:2018}, the local pairing term reads
\begin{align}
H_{\rm intra} =& \sum_{n=1,2} \frac{\Delta_{n}}{\pi a} \int dr \;  \cos [2 \theta_n(r)],
\label{Eq:H_intra_boson}
\end{align}
where $a$ is the short-distance cutoff, taken to be the transverse decay length of the hinge states.
The crossed Andreev pairing term is
\begin{align}
H_{\rm c} = \frac{2\Delta_{\rm c}}{ \pi a} \int dr \; \cos [\theta_1 (r) + \theta_2(r)] \cos [\phi_1 (r)- \phi_2(r)].
\label{Eq:H_cross_boson}
\end{align}

\begin{figure}[t]
 \includegraphics[width=\linewidth]{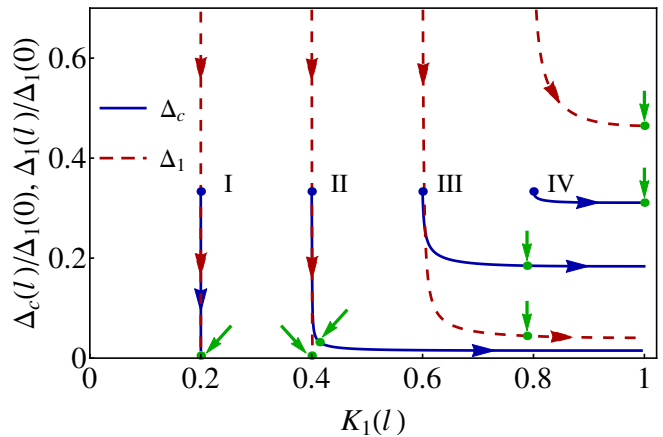}
 \caption{RG flow diagram. We take the parameters $K_{1}(0) = K_{2}(0) = 0.2$, $0.4$, $0.6$, and $0.8$, $\tilde{\Delta}_1(0) = \tilde{\Delta}_2(0) = 3\tilde{\Delta}_{\rm c}(0) = 0.03$, $a(0)=5~$nm, and $L=1~\mu$m. The crossed Andreev (local) pairing gap $\Delta_{\rm c}$ ($\Delta_{1} = \Delta_{2}$) is plotted in blue solid (red dashed) curves. The blue dots (labeled by I--IV) mark the initial points of $\Delta_{\rm c}$, and the green arrows and points specify where the RG flows stop. 
The RG flows labeled by II and III stop at the points at which the renormalized crossed Andreev pairing dominates over the local pairing ($\Delta_{\rm c} > \Delta_{1}$), indicating topological superconducting phase.
}
\label{Fig:RG_flow}
\end{figure}

Above certain interaction strength, the crossed Andreev pairing dominates and the topological criterion [see Eq.~\eqref{Eq:criterion}] is satisfied.
To show this, we derive the renormalization-group (RG) flow equations~\cite{SM} following standard procedure~\cite{Giamarchi:2003}.
To simplify the analysis, we introduce the dimensionless coupling constants,
\begin{align}
 \tilde{\Delta}_{n} (l) = \frac{\Delta_{n} (l) a(l)} { \hbar u_n },  \;\;
\tilde{\Delta}_{\rm c}(l) = \frac{\Delta_{\rm c}(l) a(l) }{ \hbar \sqrt{u_1 u_2 } }, 
\end{align}
with $l \equiv \ln [a(l)/a(0)]$.  
For given 
initial parameters (at $l=0$), we numerically propagate the RG flow equations. We stop the RG flow whenever any of the dimensionless coupling constants becomes unity. At these points we obtain the renormalized gaps and evaluate the criterion for the MKP existence. 

An example of the RG flow is in Fig.~\ref{Fig:RG_flow}, showing how the pairing gaps evolve for several starting values. 
The repulsive interaction tends to reduce both types of the pairing.
Importantly, due to the local nature of the Coulomb interaction, the suppression is stronger for the local pairing (red dashed curve) than for the crossed Andreev pairing (blue solid curve): {\it the repulsive interaction favors the nonlocal pairing.}
Consequently, even if in their initial values  the local pairing dominates over the crossed Andreev pairing [we take $\tilde{\Delta}_{\rm c}(0)/\tilde{\Delta}_1(0)=1/3$ in Fig.~\ref{Fig:RG_flow}], a sufficiently strong interaction can reverse this relation.

To prove that Eq.~\eqref{Eq:criterion} with the renormalized pairing gaps is the correct criterion,
we note that the end points of the RG flows (green arrows) are adiabatically connected to the noninteracting limit 
without closing the bulk gap. Here, the model can be refermionized into Eq.~\eqref{Eq:H_para} with renormalized pairing gaps~\cite{Gangadharaiah:2011}. 
The refermionized model can be used to justify the existence of MKPs and allows us to estimate their localization length. It is typically around 20~nm and much shorter than the hinge length $L\sim 1~\mu$m~\cite{SM}.
We thus conclude that sufficiently strong electron-electron interactions can stabilize well isolated MKPs.

 \begin{figure}[t]
 \includegraphics[width=\linewidth]{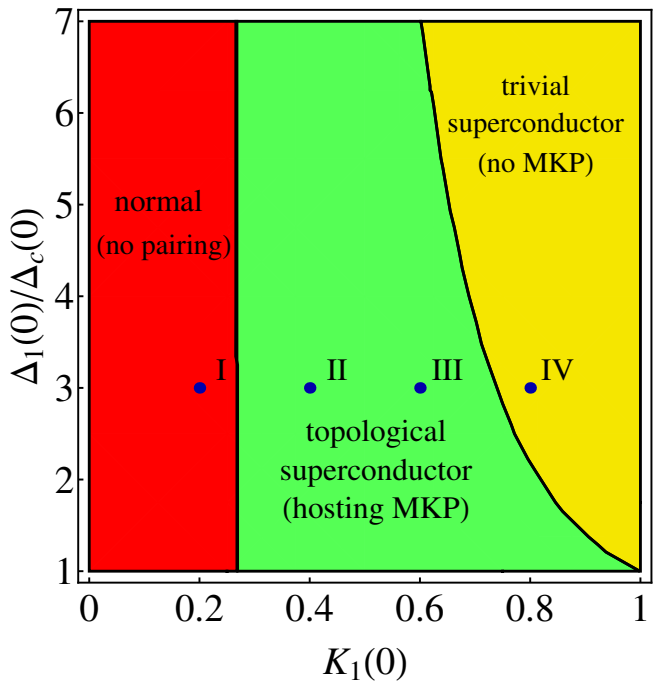}
 \caption{Phase diagram. The vertical and horizontal axes label the initial values of the gap ratio [$\Delta_{\rm 1}(0)/\Delta_{\rm c}(0)$] and interaction parameter $K_1(0)=K_2(0)$, respectively. The other parameters are the same as those in Fig.~\ref{Fig:RG_flow}. The green (yellow) region marks the phase with (without) MKPs. The corresponding RG flows to the blue dots (labeled by I--IV) are shown in Fig.~\ref{Fig:RG_flow}. In the red region, both types of the pairing gaps vanish.}
 \label{Fig:PhaseDiagram}
 \end{figure}

{\it Phase diagram.}
To investigate the stability of the MKPs in the parameter space, we repeat the above numerical procedure for $K_{1}(0) \in [0, 1]$ and $\Delta_1(0)/\Delta_{\rm c}(0) \in [1, 7]$, see Fig. \ref{Fig:PhaseDiagram}. In the phase diagram, the green (yellow) color stands for the region in which MKPs are present (absent).
If the system is noninteracting, the MKPs are absent, consistent with Ref.~\cite{Reeg:2017}. 
For $\Delta_1(0)/\Delta_{\rm c}(0) \apprge 1$, a rather weak interaction $K_{n}(0) \apprle 1$ can stabilize the MKPs. The larger 
$\Delta_1(0)/\Delta_{\rm c}(0)$ is, the stronger interaction is required to reverse the gap strengths. A very strong interaction [red region; $K_{1}(0) < 2-\sqrt{3}\approx 0.27$] destroys both types of the pairing gaps. Further, we estimate the initial values $K_1 (0) \apprle 0.6$~\cite{Maciejko:2009} and $\Delta_1(0)/\Delta_{\rm c}(0) = O(1)$~\cite{SM}, and find that they are compatible with the MKP regime in Fig.~\ref{Fig:PhaseDiagram}. 

In comparison with nonhelical nanowires~\cite{Thakurathi:2018}, our setup requires weaker interactions to induce MKPs. The difference between nonhelical and helical channels can be understood as follows. 
The effects of electron-electron interactions in nonhelical channels are ``averaged'' over charge and (noninteracting) spin sectors, and thus weakened as compared to helical channels. 
This quantitative difference indicates the advantage of helical channels, making HOTIs a promising platform for topological superconductivity without the need of magnetic fields.

We perform the same RG analysis in the standard but more involved source-term approach~\cite{Virtanen:2012,Thakurathi:2018}, in which one incorporates explicitly the inter-hinge separation $d$ and coherence length $\xi_{s}$ of the superconductor, instead of relying on knowing the initial values of $\Delta_1(0)$ and $\Delta_{\rm c}(0)$ in the effective Hamiltonian.
This analysis gives essentially the same phase diagram as in Fig.~\ref{Fig:PhaseDiagram}; see Ref.~\cite{SM}. For the parameters of bismuth hinges~\cite{Schindler:2018,Koroteev:2008,Wada:2011}, we find that moderate interactions can render a dominant crossed Andreev pairing for $d \sim 100$~nm and $\xi_{s} \sim 1~\mu$m.
With these quantitative examinations~\cite{SM}, we conclude that our setup is accessible in realistic samples.
  
{\it Discussion.}
Our work indicates that generally MKPs can be supported at the ends of a HOTI nanowire proximity coupled to a superconductor without fine-tuning.
We remark that the hinge states are known to survive even when spatial symmetries are broken by weak local perturbations due to disorder, as long as the TRS is preserved~\cite{Schindler:2018}.
As a consequence, the  MKPs proposed in this work are robust against TRS-preserving disorder.
It is also worth pointing out that, in addition to the MKPs, our setup can work as a Cooper pair splitter---a source of spatially separated spin-entangled electron pairs~\cite{Recher:2001,Recher:2002,Bena:2002,Hofstetter:2009,Sato:2010,Hofstetter:2011,Sato:2012,Schindele:2012,Das:2012b,Fulop:2014,Fulop:2015,Deacon:2015,Baba:2018}. 
Finally, we remark that detection of MKPs with the parity-controlled $2\pi$ Josephson effect, which gives distinct signatures from unpaired MBSs~\cite{Schrade:2018}, and braiding-based~\cite{Liu:2014,Gao:2016} or measurement-based~\cite{Schrade:2018b} quantum computation schemes utilizing MKPs have been widely discussed in the literature. 
Compared to setups without TRS, since here the MKPs require no magnetic fields, they are protected by a larger superconducting gap, leading to shorter localization length and longer coherence time~\cite{Schrade:2017,Schrade:2018b}. 
Our setup provides building blocks for the measurement-based structures proposed in Refs.~\cite{Landau:2016,Hoffman:2016,Karzig:2017}, which offers a route to scalable architectures for topological quantum computation.


\let\temp\addcontentsline 
\renewcommand{\addcontentsline}[3]{}

\begin{acknowledgments}
We thank R.~S.~Deacon for helpful discussion.
This work was supported financially by the JSPS Kakenhi Grant No. 16H02204, by the Swiss National Science Foundation (Switzerland), and by the NCCR QSIT. This project received funding from the European Unions Horizon 2020 research and innovation program (ERC Starting Grant, Grant Agreement No 757725). 
\end{acknowledgments}

\bibliographystyle{apsrevown}
\bibliography{MKP}

\let\addcontentsline\temp 


\clearpage
\onecolumngrid

\bigskip

\begin{center}
\large{\bf Supplemental Material to ``Majorana Kramers pairs in higher-order topological insulators''}

\fontsize{10}{12}
Chen-Hsuan Hsu$^{1}$, Peter Stano$^{1,2,3}$, Jelena Klinovaja$^{1,4}$, and Daniel Loss$^{1,4}$\\
{\it
$^{1}$RIKEN Center for Emergent Matter Science (CEMS), Wako, Saitama 351-0198, Japan \\
$^{2}$Department of Applied Physics, School of Engineering, University of Tokyo, 7-3-1 Hongo, Bunkyo-ku, Tokyo 113-8656, Japan\\
$^{3}$Institute of Physics, Slovak Academy of Sciences, 845 11 Bratislava, Slovakia\\
$^{4}$Department of Physics, University of Basel, Klingelbergstrasse 82, CH-4056 Basel, Switzerland
}
\end{center}

\twocolumngrid
\setcounter{equation}{0}
\setcounter{figure}{0}
\setcounter{table}{0}
\setcounter{page}{1}
\setcounter{NAT@ctr}{0}   
\makeatletter
\renewcommand{\theequation}{S\arabic{equation}}
\renewcommand{\thefigure}{S\arabic{figure}}
\renewcommand{\bibnumfmt}[1]{[S#1]}
\renewcommand{\citenumfont}[1]{S#1}

\tableofcontents

\subsection{I. Pairing processes and decoupling of the third hinge}

\begin{figure}[th]
 \includegraphics[width=\linewidth]{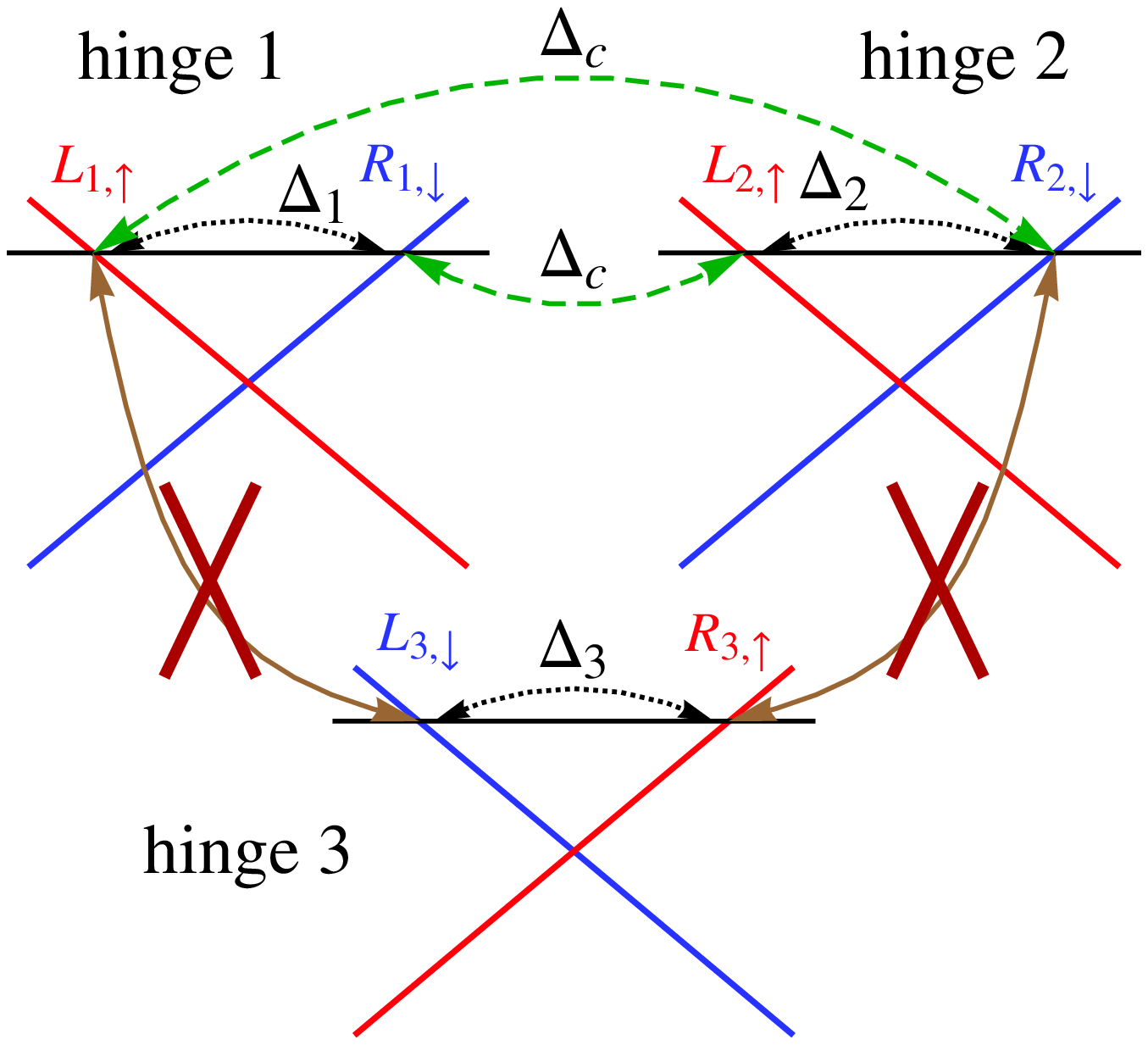}
 \caption{Schematics of the pairing processes in momentum space. The hinges 1 and 2 carry the same helicity (a given spin state in both of the hinges move in the same direction) while the hinge 3 has the opposite helicity. The local pairing (black dotted arrows, $\Delta_{n}$) occurs within the hinges. 
 The crossed Andreev pairing 
 (green dashed arrows, $\Delta_{\rm c}$) is allowed only between the hinges 1 and 2, and forbidden between the other pairs due to the momentum mismatch.}
 \label{Fig:momentum}
 \end{figure}

In this Section we discuss the pairing processes within/between the three helical hinges, and show that the third, middle parallel, hinge is trivially gapped and decoupled from the other hinges (1 and 2). We describe the three hinge states by the fields,
\begin{subequations}
\begin{align}
\psi_n (r) =& \psi_{n,\downarrow} (r) + \psi_{n,\uparrow} (r)  \;\; {\rm for\;} n \in \{1,2,3\}, \\
 \psi_{n,\uparrow} (r)  =& L_{n, \uparrow}(r) e^{-i k_F r}  \;\; {\rm for\;} n \in \{1,2\}, \\
 \psi_{n,\downarrow} (r) =& R_{n, \downarrow} (r) e^{i k_F r}   \;\; {\rm for\;} n \in \{1,2\}, \\
 \psi_{3,\uparrow} (r)  =&  R_{3, \uparrow} (r) e^{i k_F r} , \\
\psi_{3,\downarrow} (r) =&  L_{3, \downarrow}(r)  e^{-i k_F r}, 
\end{align} 
\end{subequations}
with the coordinate $r$ along the hinge, the Fermi wave number $k_F$, and the slowly varying right- and left-moving fields $R_{n, \sigma}$ and $L_{n, \sigma}$ (the spin index $\sigma \in \{\uparrow, \downarrow\}$), respectively. 
A Bardeen--Cooper--Schrieffer (BCS) type pairing term between the hinges $n$ and $n'$ is in the form of
\begin{align}
 \propto [\psi_{n,\downarrow}^{\dagger} (-k) \psi_{n',\uparrow}^{\dagger} (k) + {\rm H.c.}],
\end{align}
 with the momentum $k$, which puts constraints on the momenta and spins of the allowed pairs. Namely, for the electrons to be paired, they must possess zero total momentum and opposite spins. The pairing processes within/between the three helical hinges are sketched in Fig.~\ref{Fig:momentum}. The local pairing terms ($n=n'$) for hinges 1 and 2 are given by Eq.~\eqref{Eq:H_intra} in the main text, and, for hinge 3, it is given by
\begin{eqnarray}
H_{\rm 3}  &=& \int dr \; \Big[ \frac{\Delta_{3}}{2} (R_{3,\uparrow}^\dagger L_{3,\downarrow}^\dagger - L_{3,\downarrow}^\dagger R_{3,\uparrow}^\dagger) + {\rm H.c.} \Big],
\label{Eq:H_intra3}
\end{eqnarray}
with the local pairing gap $\Delta_{3}$. These terms are indicated by the black dotted arrows in Fig.~\ref{Fig:momentum}.

On the other hand, the nonlocal (crossed Andreev) pairing term  ($n \neq n'$) requires more caution. 
Due to the identical chemical potential of the hinges, the pairing between the parahelical hinges (1 and 2) is allowed, which is described by the green dashed arrows in Fig.~\ref{Fig:momentum} [see Eq.~\eqref{Eq:H_cross} in the main text], whereas the corresponding terms between the hinges 3 and $n \in \{1,2\}$ are given by 
\begin{align}
\int dr \; \Big\{ \big( R_{n,\downarrow}^\dagger R_{3,\uparrow}^\dagger e^{-2i k_F r} + 
L_{3,\downarrow}^\dagger L_{n,\uparrow}^\dagger e^{2i k_F r} \big) + {\rm H.c.} \Big\}. 
\label{Eq:H_cross3}
\end{align}
 The integral in Eq.~\eqref{Eq:H_cross3} contains a fast oscillating integrand, and vanishes for $k_F L \gg 1$. 
As a result, the crossed Andreev pairing between the third hinge and the others is suppressed due to the momentum mismatch, so we can describe the two subsystems separately. To be specific, we describe the whole system of the three hinges by the total Hamiltonian, $H_{\rm tot} = H_{\rm  } + H_{\rm 3}$, which is block-diagonalized into two parts, 
\begin{subequations}
\begin{eqnarray}
H_{\rm  } &=& \frac{1}{2} \int dr \; \Psi^{\dagger} (r) \mathcal{H}_{\rm  } (r) \Psi (r), \\
H_{\rm 3} &=& \frac{1}{2} \int dr \; \Psi_{3}^{\dagger} (r) \mathcal{H}_{\rm 3} (r) \Psi_{3} (r),
\end{eqnarray}
\end{subequations}
with the basis
\begin{subequations} 
\begin{align}
\Psi =& (R_{1,\downarrow}, L_{1,\uparrow}, R_{2,\downarrow}, L_{2,\uparrow}, R_{1,\downarrow}^\dagger, L_{1,\uparrow}^\dagger, R_{2,\downarrow}^\dagger, L_{2,\uparrow}^\dagger)^{\rm T}, \\
\Psi_{3} =& (R_{3,\uparrow}, L_{3,\downarrow},  R_{3,\uparrow}^\dagger, L_{3,\downarrow}^\dagger)^{\rm T}.
\end{align} 
\end{subequations}
The Hamiltonian density $\mathcal H_{\rm  }$ in the subspace of hinges 1 and 2 is given in Eq.~\eqref{Eq:H_para} in the main text. In the subspace of hinge 3, we have 
\begin{eqnarray}
\mathcal{H}_{\rm 3}(r) &=& -i \hbar v_F \eta^0 \sigma^z \partial_r  -\Delta_{3} \eta^y \sigma^y,
\end{eqnarray}
so the third hinge is trivially gapped by the local pairing, and decoupled from the others. This allows us to focus on the two hinges that are parahelical to each other. We therefore present an effective model $\mathcal H_{\rm  }$ for the two parahelical hinges of interest (1 and 2) in the main text.

In the above, we take the same Fermi wave number for the parahelical hinges 1 and 2 based on the following considerations.  
First, the chemical potential is uniform along the hinges, since they are all connected. Second, the bismuth crystal structure respects the three-fold rotational symmetry around the (111) axis~\citesupp{Schindler:2018_S}. The two parahelical hinges, which are related to each other by such a rotation, are therefore characterized by the same energy band, as implied by the $C_3$ rotational symmetry of the underlying crystal structure; see
Ref.~\citesupp{Schindler:2018_S} for the degenerate hinge-state energy bands calculated from their tight-binding model.

Local disorder may, however, alter the energy bands of the hinge states, and therefore modify the Fermi wave numbers.
As a result, a difference between the Fermi wave numbers of the two hinges will reduce the crossed Andreev pairing. Such an effect can be straightforwardly incorporated in our model, by starting with a larger ratio of $\Delta_{1}/\Delta_{\rm c}$ in the renormalization-group (RG) analysis. Assuming the perturbations are small (weak disorder) so that the suppression is not complete, the Majorana Kramers pairs will still be stabilized (even though stronger electron-electron interactions are then required), and our conclusions remain qualitatively unchanged.

\subsection{II. Wave functions of the Majorana Kramers pair}
In this Section we give the bulk spectrum of the single-particle Hamiltonian $\mathcal{H}(r)$ in Eq.~\eqref{Eq:H_para}, and the corresponding wave functions of the Majorana Kramers pair localized at the boundary $r=0$.
Upon replacement $-i \partial_r \rightarrow k$ in $\mathcal{H}(r)$ and diagonalization, we find the two-fold degenerate bulk spectrum 
\begin{align}
E_{\rm bulk}^{(\pm,\pm)} (k) =& \pm \Big[ \hbar^2 v_F^2 k^2 + \Big( \Delta_{+} \pm \sqrt{\Delta_{-}^2 + \Delta_{\rm c}^2 } \Big)^2 \Big]^{1/2},
\end{align}
which has a gap at $k=0$. To be specific, we define the bulk gap as
\begin{eqnarray}
\Delta_{\rm b} &\equiv & E_{\rm bulk}^{(+,-)} (k=0) - E_{\rm bulk}^{(-,-)} (k=0) \nonumber \\
&=& 2 \Big( \Delta_{+} - \sqrt{\Delta_{-}^2 + \Delta_{\rm c}^2 } \Big).
\label{Eq:Delta_b}
\end{eqnarray}
Assuming $\Delta_{1}, ~\Delta_{2} >0$, the sign of $\Delta_{\rm b}$ is given by the sign of
\begin{align}
\Delta_{1} \Delta_{2} - \Delta_{\rm c}^2,
\end{align}
indicating that the bulk gap $\Delta_{\rm b}$ changes it sign when the local and crossed Andreev pairings reverse their relative strengths.
 
We remark that, for $\Delta_1$ and $\Delta_2$ with general signs, $\Delta_{\rm b}$ can reverse its sign without involving the crossed Andreev pairing. It can be demonstrated by setting $\Delta_{\rm c}=0$ in Eq.~\eqref{Eq:Delta_b}. In this case, $\Delta_{\rm b}$ becomes negative when the signs of $\Delta_1$ and $\Delta_2$ are opposite ($\pi$-junction).
In our setup, the phase of $\Delta_1$ and $\Delta_2$ is the same, as they both stem from a single parent superconductor. An interesting scenario of a $\pi$-junction\citesupp{Fu:2008_S,Wong:2012_S,Keselman:2013_S,Zhang:2013_S,Haim:2014_S,Schrade:2015_S,Hoffman:2016b_S} would require a different setup. 
In the main text, we therefore focus on the case where $\Delta_1$ and $\Delta_2$ are both positive.

We now turn to the wave functions of the Majorana Kramers pair. 
When $\Delta_{\rm c}^2 > \Delta_{1} \Delta_{2}$, we find that a Kramers pair of Majorana bound states emerges at the boundary $r=0$, with the wave functions $\Phi_{\rm MF, 1}$ and $\Phi_{\rm MF, 2}$. In the basis of $\Psi = (R_1, L_1, R_2, L_2, R_1^\dagger, L_1^\dagger, R_2^\dagger, L_2^\dagger)^{\rm T}$,
we have $\Phi_{\rm MF, 1}(r) = \Phi_{>}(r)\Theta(r) + \Phi_{<}(r)\Theta(-r)$, with the step function $\Theta(r)$ and 
\begin{subequations}
\begin{align}
	\Phi_{>}(r) =& e^{-\kappa r} \times  (i \eta, - \eta, -i, 1, -i \eta, - \eta, i, 1)^{\rm T} ,\\
	\Phi_{<}(r) =& (i \eta e^{\kappa_1 r}, - \eta e^{\kappa_1 r}, -i e^{\kappa_2 r}, e^{\kappa_2 r},  \nonumber\\
	& \hspace{0pt} -i \eta e^{\kappa_1 r}, - \eta e^{\kappa_1 r}, i e^{\kappa_2 r},  e^{\kappa_2 r})^{\rm T} ,
\end{align}
\end{subequations}
where the normalization constants of $\Phi_{>}(r)$ and $\Phi_{<}(r)$ were omitted, and 
\begin{subequations}
\begin{align}
	\eta = & \frac{ \sqrt{\Delta_{-}^2 + \Delta_{\rm c}^2} - \Delta_{-} }{\Delta_{\rm c}},	\\
	\kappa =& \frac{ \sqrt{\Delta_{-}^2 + \Delta_{\rm c}^2} - \Delta_{+} }{\hbar v_F}, \\
	\kappa_n =& \frac{\Delta_n }{\hbar v_F} \;\; {\rm for\;} n \in \{1,2\}. 
\end{align}
\end{subequations}
The localization length of the Majorana bound states is thus given by 
\begin{align}
\xi_{\rm loc} = \frac{1}{\min (\kappa, \kappa_{1}, \kappa_{2})},
\label{Eq:xi_loc}
\end{align}
which is estimated in Sec.~IV. 
The second Majorana wave function is related to its Kramers partner by $\Phi_{\rm MF, 2} = {\mathcal T} \Phi_{\rm MF, 1}$ with the time-reversal operator $\mathcal{T}=i \sigma^y\mathcal{K}$ and the complex conjugate operator $\mathcal{K}$. We note that the two Majorana wave functions also satisfy the relation $\Phi_{\rm MF, 1} = - {\mathcal T} \Phi_{\rm MF, 2}$, such that ${\mathcal T}^2 = -1$, as required for spin-$1/2$ particles. 
One can check that $\Phi_{\rm MF, 1}$ and $\Phi_{\rm MF, 2}$ are orthogonal, as guaranteed by the Kramers degeneracy theorem. Therefore, even though they are not spatially separated, they do not hybridize into an ordinary fermion as long as time-reversal symmetry is preserved.

\subsection{III. Effects of the intra-/inter-hinge coupling and the spin-orbit interactions}
In this Section we discuss the effects of the intra-hinge and inter-hinge couplings, as well as the spin-orbit interactions.
In addition to the pairing processes discussed in the main text, the single-particle intra-hinge (inter-hinge) coupling can also arise from the second-order spin-conserving (co-)tunneling processes within (between) hinges 1 and 2. We note that the effects of similar processes have been discussed in nonhelical systems~\citesupp{Reeg:2017_S,Schrade:2017_S} and in the helical edge channels of two-dimensional topological insulators~\citesupp{Klinovaja:2014_S,Klinovaja:2015_S}.

To proceed, we incorporate the following terms
 \begin{align}
H_{\delta \mu} &= - \delta \mu \sum_{n=1,2} \sum_{\sigma=\uparrow,\downarrow} \int dr \; \psi_{n,\sigma}^\dagger(r) \psi_{n,\sigma}(r) ,\\
H_{\Gamma} &= -\Gamma \sum_{\sigma=\uparrow,\downarrow} \int dr \; \big[\psi_{1,\sigma}^\dagger(r) \psi_{2,\sigma}(r) + {\rm H.c.} \big],
\label{Eq:H_single-coupling1}
\end{align}
into the single-particle Hamiltonian given in the main text. To be explicit, we replace the Hamiltonian density [Eq.~\eqref{Eq:H_para} in the main text] with
 \begin{align}
\mathcal{H} \rightarrow \mathcal{H} -\delta \mu \, \eta^z \tau^0 \sigma^0 -\Gamma \, \eta^z \tau^x \sigma^0
\label{Eq:H_single-coupling2}
\end{align}
with the Pauli matrices defined as in the main text. The effective couplings $\delta \mu$ and $\Gamma$ are of the second order in the tunnel coupling between the hinges and the superconductor. The derivations of the explicit forms of $\delta \mu$ and $\Gamma$ can be nontrivial~\citesupp{Reeg:2017_S,Schrade:2017_S}. Instead of aiming at the  derivation, we treat them as effective parameters, and show that they do not affect the topological criterion.

The bulk spectrum of the full Hamiltonian [with the Hamiltonian density in Eq.~\eqref{Eq:H_single-coupling2}] is given by $ \pm  E_{\rm b}^{(\eta,\pm)} (k)$, with
\begin{align}
E_{\rm b}^{(\eta,\pm)} (k) =& \Big[ \xi_{\eta}^2(k) + \Delta_{1}^2 + \Delta_{\rm c}^2 + \Gamma^2 \nonumber \\
& \hspace{16pt} \pm 2 \big| \xi_{\eta}(k) \Gamma + \eta \Delta_{1} \Delta_{\rm c} \big| \Big]^{1/2},
\end{align}
where we define $\xi_{\eta}(k) \equiv \hbar v_F k  + \eta \delta \mu$ with $\eta \in \{+,-\}$ and set $\Delta_1=\Delta_2$ for simplicity. We note that the bulk spectrum remains two-fold degenerate at the time-reversal-invariant momentum $k=0$. From the above equation, we see that, for general $\Gamma$ and $\delta \mu$, the band-touching points shift from $k=0$ to the points associated with $\xi_{\eta}^2(k) = \Gamma^2$. Important for us, however, the system gap remains to be $|\Delta_{\rm c} - \Delta_{1}|$ (assuming $\Delta_{1},~\Delta_{\rm c}>0$), indicating the same gap closing point and therefore the same phase transition point in the parameter space
\begin{align}
\Delta_{\rm c} = \Delta_{1},
\end{align}
as in the limit of $\Gamma = \delta \mu = 0$. 
As a consequence, the nonzero $\Gamma$ and $\delta \mu$ do not lead to the system gap closing or reopening, so that the Majorana Kramers pairs in the $\Delta_{\rm c} > \Delta_{1}$ regime are robust against the intra-hinge and inter-hinge couplings as long as the pairing gaps $\Delta_n$ are not drastically different. 

Furthermore, it was shown in Ref.~\citesupp{Klinovaja:2014_S} that when the electron-electron interactions are taken into account, the crossed Andreev pairing is the most relevant term in the RG sense (justified by the scaling dimension). As a result, the additional tunneling processes are less important than the crossed Andreev pairing, and can be safely neglected in interacting systems. To conclude, while the intra-hinge and inter-hinge couplings alter the bulk spectrum, they do not destabilize the Majorana Kramers pairs.

Finally, it may be interesting to examine the effects of the spin-orbit interactions arising from, say, substrates or gate-induced electric field. In principle, these Rashba-type spin-orbit interactions may affect the spins of the hinge states, similarly to edges states of two-dimensional topological insulators~\citesupp{Schmidt:2012_S}. While the detailed analysis on their effects is beyond the scope of this work, we make two remarks here. First, as mentioned in the main text, due to its self-tuned nature, our setup does not require local gates, thereby minimizing the Rashba-type spin-orbit interactions. Second, since the spin-orbit interactions respect time-reversal symmetry, there are still Kramers partners of the hinge states. Consequently, the spin states of the left- and right-movers in a given hinge remain strictly orthogonal, so the pairing processes are not affected by the spin-orbit interactions. Concluding, the presence of either the intra-/inter-hinge coupling or the spin-orbit interactions does not alter our conclusions.

\subsection{IV. RG analysis from the effective Hamiltonian method}
In this Section we outline our RG analysis using the effective Hamiltonian method.
First, we introduce the boson fields $\theta_n$ and $\phi_n$ used to bosonize the Hamiltonian in an interacting system. They are related to the fermion fields $R_{n}$ and $L_{n}$ through (with the hinge index $n \in \{ 1, 2 \}$)~\citesupp{Hsu:2017_S,Hsu:2018_S}
\begin{subequations}
\label{Eq:bosonization}
\begin{eqnarray}
R_{n} (r) &=& \frac{U_{R}}{\sqrt{2\pi a}} e^{i[-\phi_{n}(r) + \theta_{n}(r)]}, \\
L_{n} (r) &=& \frac{U_{L}}{\sqrt{2\pi a}} e^{i[\phi_{n}(r) + \theta_{n}(r)]},
\end{eqnarray}
\end{subequations} 
where $U_{R}$ and $U_{L}$ are the Klein factors, and $a$ is the short-distance cutoff, taken to be the transverse decay length of the hinge states. The formulas \eqref{Eq:bosonization} are used to derive Eqs.~\eqref{Eq:H_el}, \eqref{Eq:H_intra_boson}, and \eqref{Eq:H_cross_boson} in the main text.

Next, we present the RG flow equations derived from the effective Hamiltonian method [see Eqs.~\eqref{Eq:H_el}, \eqref{Eq:H_intra_boson}, and \eqref{Eq:H_cross_boson} in the main text]. 
Since Eq.~\eqref{Eq:H_cross_boson} contains the fields $\phi_n$ while Eq.~\eqref{Eq:H_intra_boson} contains their conjugate fields $\theta_n$, the two types of pairing processes compete with each other~\citesupp{Giamarchi:2003_S}. We then expect that their relative strength varies with the interaction strength, as we demonstrate below.

To this end, we construct the RG flow equations by computing the correlation function, and changing the cutoff $a \rightarrow a(l)=a(0) e^{l}$ with the dimensionless scale $l = \ln [a(l)/a(0) ]$~\citesupp{Giamarchi:2003_S}. We obtain 
\begin{subequations}
\label{Eq:RG_EffH}
\begin{eqnarray}
\frac{d \tilde{\Delta}_1 (l)}{d l} &=& \left[2 - \frac{1}{K_1(l)} \right] \tilde{\Delta}_1 (l), \\
\frac{d \tilde{\Delta}_2 (l)}{d l} &=& \left[2 - \frac{1}{K_2(l)} \right] \tilde{\Delta}_2 (l), \\
\frac{d \tilde{\Delta}_{\rm c} (l)}{d l} &=& \left[2 - \frac{1}{4} \Big( K_1(l) + K_2 (l) + \frac{1}{K_1(l)}  +\frac{1}{K_2(l)} \Big) \right] \nonumber\\
&& \hspace{0.0in} \times \tilde{\Delta}_{\rm c} (l), \\
\frac{d K_1 (l)}{d l} &=& \tilde{\Delta}_1^2 (l) + \frac{1}{2} \left[1 - K_1^2(l) \right] \tilde{\Delta}_{\rm c}^2 (l) ,\\
\frac{d K_2 (l)}{d l} &=& \tilde{\Delta}_2^2 (l) + \frac{1}{2} \left[1 - K_2^2(l) \right] \tilde{\Delta}_{\rm c}^2 (l),
\end{eqnarray}
\end{subequations}
 where the dimensionless coupling constants are given by
\begin{subequations}
\label{Eq:dimensionless}
\begin{eqnarray}
&\tilde{\Delta}_1(l) = \frac{\Delta_1(l) a(l)}{\hbar u_1 }, \\
&\tilde{\Delta}_2(l) = \frac{\Delta_2(l) a(l)}{\hbar u_2 }, \\
&\tilde{\Delta}_{\rm c}(l) =  \frac{\Delta_{\rm c}(l)  a(l)}{\hbar \sqrt{u_1 u_2} }. 
\end{eqnarray}
\end{subequations}
Several remarks on Eqs.~\eqref{Eq:RG_EffH} are in order. First, both types of dimensionful pairing gaps $\Delta_{n}$ and $\Delta_{\rm c}$ are suppressed by a repulsive interaction $K_{n}(0)<1$. Second, due to the different scaling dimensions of the cosine terms in Eqs.~\eqref{Eq:H_intra_boson}--\eqref{Eq:H_cross_boson}, the local pairing gap is suppressed more significantly than the crossed Andreev pairing. Therefore, the repulsive interaction favors the nonlocal pairing process, similarly as in nonhelical nanowires~\citesupp{Thakurathi:2018_S}. Third, from Eqs.~\eqref{Eq:RG_EffH}, we see that, for $K_{1}(0) = K_{2}(0) < 2-\sqrt{3}\approx 0.27$, both $\tilde{\Delta}_{1,2}$ and $\tilde{\Delta}_{\rm c}$ are irrelevant in the RG sense. Namely, both of the gaps flow to zero, so the hinge states remain non-superconducting in the presence of a very strong electron-electron interaction.

The RG flow equations Eqs.~\eqref{Eq:RG_EffH} are numerically solved using the initial conditions, 
\begin{subequations}
\label{Eq:initial}
\begin{eqnarray}
\tilde{\Delta}_1 (0) &\equiv& \tilde{\Delta}_1 (l=0)  = \tilde{\Delta}_2 (l=0) , \\
\tilde{\Delta}_{\rm c}(0) &\equiv& \tilde{\Delta}_{\rm c} (l=0),  \\
K_1 (0) &\equiv&  K_1 (l=0) =  K_2 (l=0).
\end{eqnarray}
\end{subequations}
The RG flow is stopped whenever any of the dimensionless coupling constants becomes unity, including also the interaction parameters $K_n$ and the dimensionless system size $a(l)/L$ with the hinge length $L$. 
The main results of the numerical calculation are presented in Figs.~\ref{Fig:RG_flow} and \ref{Fig:PhaseDiagram}, and discussed in the main text. Moreover, the renormalized values of the pairing gaps allow us to obtain the localization length. Using Eq.~\eqref{Eq:xi_loc} and the initial parameters $K_{1}(0) = K_{2}(0) = 0.6$--$0.7$, $\tilde{\Delta}_1(0) = \tilde{\Delta}_2(0) = 0.03$, $\tilde{\Delta}_{\rm c}(0) = 0.01$, and $a(0)=5~$nm, we get $\xi_{\rm loc}=19$--21~nm, which is much shorter than the hinge length $L \sim O(\mu$m), and justifies that the Majorana bound states located at $r=0$ and $r=L$ do not overlap.

In comparison with nonhelical (spin-degenerate) nanowires~\citesupp{Thakurathi:2018_S}, our setup requires weaker interactions for Majorana Kramers pairs. This quantitative difference between the nonhelical and helical channels can be understood as discussed in Refs.~\citesupp{Hsu:2017_S,Hsu:2018_S}. 
Namely, in a nonhelical system, the effects of the electron-electron interactions are ``averaged'' over the charge and (noninteracting) spin sectors, and thus weakened compared to a helical one. 
Consequently, it requires stronger interactions in the charge sector of a nonhelical channel to drive the system into the regime hosting Majorana Kramers pairs. It suggests that the helical hinge states of higher-order topological insulators offer a promising platform for topological superconductivity without magnetic fields.

\subsection{V. Source-term approach using a microscopic model}
In this Section we present the RG flow equations and phase diagram obtained from the source-term approach~\citesupp{Virtanen:2012_S,Thakurathi:2018_S}, supplementary to the effective Hamiltonian method presented in the main text.
We assume a weak tunnel coupling between the hinge states and a proximity BCS superconductor, described by the tunnel Hamiltonian
\begin{align}
H_{\rm T} &=& \sum_{n=1,2} \int dr d{\bf R} \; \Big\{ t^{\prime}_{n} (r,{\bf R}) \Big[R_n^{\dagger}(r) \psi_{{\rm s}, \downarrow}({\bf R})  \nonumber \\
&& \hspace{0.85in} + L_{n}^{\dagger} (r) \psi_{{\rm s},\uparrow} ({\bf R}) \Big] + {\rm H.c.} \Big\}. 
\label{Eq:H_T}
\end{align}
Here, ${\bf R}$ is the three-dimensional coordinate in the bulk of the superconductor, and $\psi_{{\rm s}, \sigma}$ is the annihilation operator with spin $\sigma$ in the superconductor. We take the tunnel amplitude $t^{\prime}_{n}$ to be in the form of the three-dimensional delta function,
\begin{subequations}
\begin{eqnarray}
t^{\prime}_{n} (r,{\bf R}) &\equiv& t_{n} \delta (R_z-r) \delta (R_x- d_n) \delta (R_y),
\end{eqnarray}
\end{subequations}
with $d_{1} = d/2$ and $d_{2} = -d/2$, where $d$ is the distance between the two hinges. 
The BCS Hamiltonian describing the superconductor is given by 
\begin{eqnarray}
H_{\rm BCS} &=& \sum_{{\bf k}, \sigma = \uparrow, \downarrow} 
 \frac{\hbar^2 ( k^2  -  k_{F {\rm s}}^2  ) }{2 m_e}
 \psi_{{\rm s}, \sigma}^{\dagger}({\bf k}) \psi_{{\rm s}, \sigma}({\bf k}) \nonumber\\
&& + \Delta_{\rm s} \sum_{{\bf k}} \psi_{{\rm s}, \uparrow}({\bf k}) \psi_{{\rm s}, \downarrow}({\bf -k}) + {\rm H.c.}, 
\label{Eq:H_BCS}
\end{eqnarray}
with the electron mass $m_e$, the BCS pairing gap $\Delta_{\rm s}$, and the Fermi wave number $k_{F, {\rm s}}$ of the superconductor in its normal phase.

To proceed, we first integrate out the field $\psi_{{\rm s}, \sigma}$ in $H_{\rm BCS}+ H_{\rm T}$ to obtain the effective action~\citesupp{Reeg:2017_S}, and then construct the RG flow equations following Refs.~\citesupp{Giamarchi:2003_S,Virtanen:2012_S,Thakurathi:2018_S}. The result is summarized as follows,
\begin{subequations}
\label{Eq:RG_SourceTerm}
\begin{eqnarray}
\frac{d \tilde{t}_1 (l)}{d l} &=& \left[2 - \frac{1}{4} \Big ( K_1(l) +\frac{1}{K_1(l)} \Big) \right] \tilde{t}_1 (l), \\
\frac{d \tilde{t}_2 (l)}{d l} &=& \left[2 - \frac{1}{4} \Big ( K_2(l) +\frac{1}{K_2(l)} \Big) \right] \tilde{t}_2 (l), \\
\frac{d \tilde{\Delta}_1 (l)}{d l} &=& \left[2 - \frac{1}{K_1(l)} \right] \tilde{\Delta}_1 (l) + S_{1}(l) \tilde{t}_1^2 (l), \\
\frac{d \tilde{\Delta}_2 (l)}{d l} &=& \left[2 - \frac{1}{K_2(l)} \right] \tilde{\Delta}_2 (l) + S_{2}(l) \tilde{t}_2^2 (l), \\
\frac{d \tilde{\Delta}_c (l)}{d l} &=& \left[2 - \frac{1}{4} \Big ( K_1(l) + K_2 (l) +\frac{1}{K_1(l)} +\frac{1}{K_2(l)} \Big) \right] \nonumber\\
&& \hspace{0.0in} \times \tilde{\Delta}_c (l) 
+ S_{c}(l) \tilde{t}_1 (l) \tilde{t}_2 (l), \\
\frac{d K_1 (l)}{d l} &=& \tilde{\Delta}_1^2 (l) + \frac{1}{2} \left[1 - K_1^2(l) \right] \tilde{\Delta}_c^2 (l), \\
\frac{d K_2 (l)}{d l} &=& \tilde{\Delta}_2^2 (l) + \frac{1}{2} \left[1 - K_2^2(l) \right] \tilde{\Delta}_c^2 (l), 
\end{eqnarray}
\end{subequations}
where the dimensionless coupling constants for the pairing gaps are given in Eqs.~\eqref{Eq:dimensionless}, and the dimensionless tunnel amplitude is given by
\begin{eqnarray}
\tilde{t}_n(l) &=& t_n(l) \sqrt{ \frac{a^3(l)}{ \hbar^2 u_n^2 \xi_{\rm s}^2 L }},
\end{eqnarray}
with the hinge length $L$ and the superconducting coherence length $\xi_{\rm s}$. 
 The coefficients of the source terms are given by
\begin{subequations}
\label{Eq:SourceTerm}
\begin{eqnarray}
 S_{n}(l)  &=& \frac{m_e v_{F {\rm s}}^2 L}{2\pi \Delta_{\rm s} a(l)} K_{0} \Big( \frac{\Delta_{\rm s} a (l)}{ \hbar u_n} \Big),\\
 S_{\rm c}(l)  &=& \frac{m_e v_{F {\rm s}}^2 L}{2\pi \Delta_{\rm s} d} e^{-d/\xi_s} \left| \sin (k_{F {\rm s}} d) \right| \nonumber \\
&& \times I_{0} \Big( \frac{\Delta_{\rm s}   a (l) }{2 \hbar \sqrt{u_1 u_2} } \Big) K_{0} \Big( \frac{\Delta_{\rm s} a(l) }{2 \hbar \sqrt{u_1 u_2} } \Big),
\end{eqnarray}
\end{subequations}
with the modified Bessel function of the first- (second-)kind, $I_0(x)$ [$K_{0}(x)$]. The behavior of Eqs.~\eqref{Eq:SourceTerm} is discussed in Ref.~\citesupp{Thakurathi:2018_S}, so not repeated here. Rather, we point out that the source terms naturally include the effects of the coherence length and the distance $d$ between the hinges. Namely, due to the factor $e^{-d/\xi_s} |\sin (k_{F {\rm s}} d) |\xi_s/d$, the strength of the crossed Andreev pairing is reduced when the distance $d$ increases~\citesupp{Reeg:2017_S}. 
In contrast, in the effective Hamiltonian model, such a dependence is not explicitly included, and has to be incorporated by setting an initial ratio $\Delta_{\rm c}(0)/\Delta_{1}(0)<1$. 
Nonetheless, as demonstrated below, our numerical calculation show that such a reduction is modest for a small ratio of $d/\xi_s$, so that the crossed Andreev pairing can eventually dominate over the local pairing in the presence of interactions.

 \begin{figure}[t]
 \includegraphics[width=\linewidth]{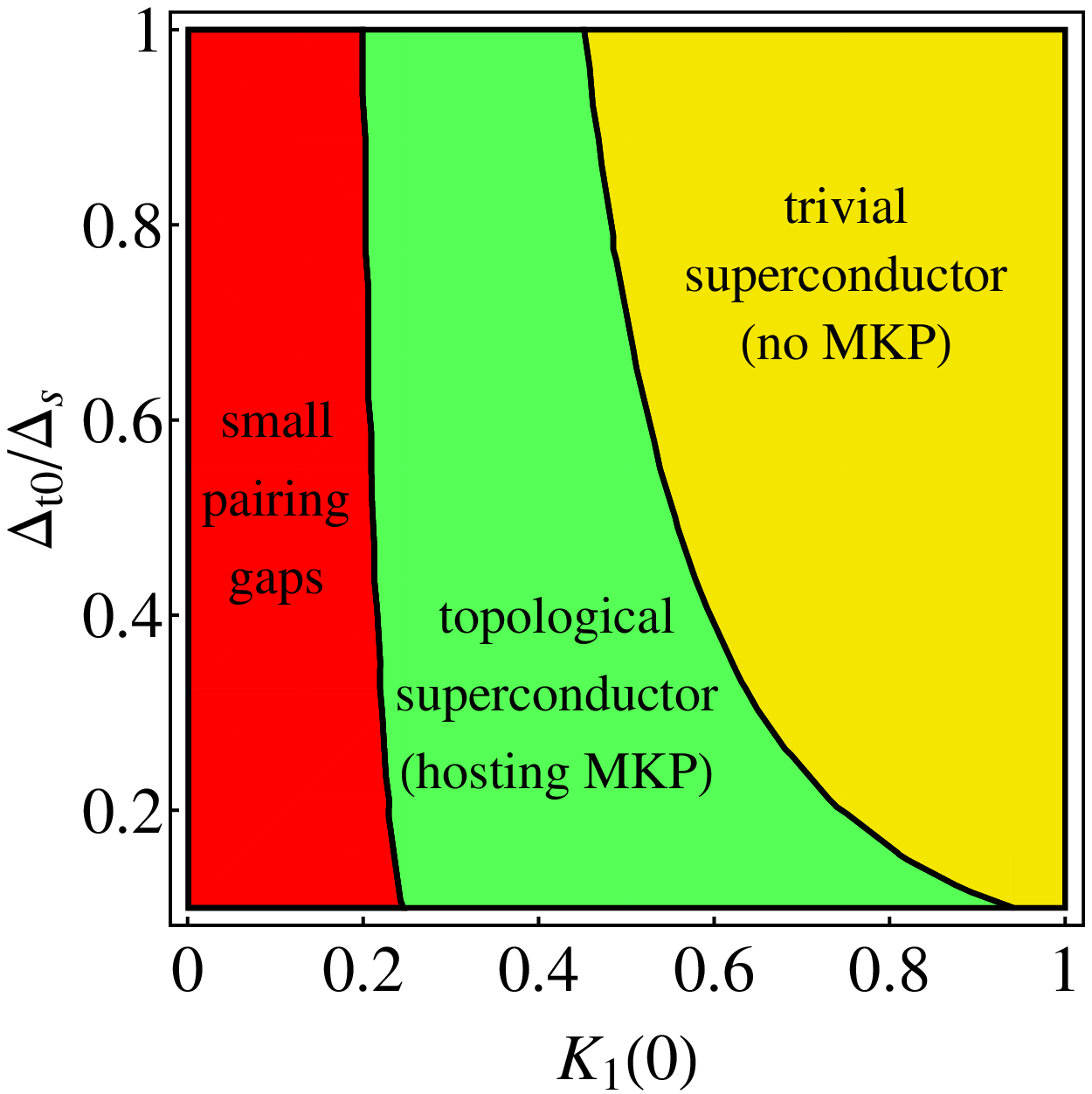}
 \caption{A phase diagram obtained from the source-term approach, by solving the RG flow equations in Eqs.~\eqref{Eq:RG_SourceTerm}. 
Here we define $\Delta_{t0} \equiv \tilde{t}_{1}^2(0) S_{1}(0) \hbar u_1/a(0)$, and take the following parameters: $u_1=u_2= 10^5$~m/s, $a(0)=5~$nm, $L=1~\mu$m, $\Delta_s=0.35$~meV, $k_{F {\rm s}}=10^{10}$~m$^{-1}$, and $d=50~$nm. As in Fig.~\ref{Fig:PhaseDiagram} in the main text, the green (yellow) region specifies the phase with (without) a Majorana Kramers pair. In the red region, the renormalized values of both $\tilde{\Delta}_{\rm c}$ and $\tilde{\Delta}_{1}$ are less than 0.1.}
 \label{Fig:PhaseDiagram_ST}
 \end{figure}

The RG flow equations in Eqs.~\eqref{Eq:RG_SourceTerm} are numerically solved for the following initial parameter values
\begin{subequations}
\label{Eq:initial_ST} 
\begin{eqnarray}
\tilde{t}_{1}(0) &\equiv& \tilde{t}_1 (l=0)  = \tilde{t}_2 (l=0), \\
K_1 (0) &\equiv&  K_1 (l=0) =  K_2 (l=0) ,\\
 S_{1}(0)  &\equiv&  S_{1}(l=0) =  S_{2}(l=0) ,\\
 S_{\rm c}(0)  &\equiv& S_{\rm c}(l=0), \\
 \tilde{\Delta}_1 (l=0) & = & \tilde{\Delta}_2 (l=0) =0, \\
\tilde{\Delta}_{\rm c}(0) &=& \tilde{\Delta}_{\rm c} (l=0) = 0.
\end{eqnarray}
\end{subequations}
This means that the initial values of the pairing gaps are set to zero. Under the RG flow, the pairing gaps are induced by the source terms arising from the tunnel Hamiltonian [see Eq.~\eqref{Eq:H_T}]. 
We obtain the phase diagram by solving Eqs.~\eqref{Eq:RG_SourceTerm} with the initial parameters [see Eqs.~\eqref{Eq:initial_ST}] in the regime $K_{1}(0) \in [0,1]$ and $\Delta_{t0}/\Delta_{\rm s} \in [0.1, 1]$, as displayed in Fig.~\ref{Fig:PhaseDiagram_ST}.   
The result is qualitatively similar to the one obtained from the effective Hamiltonian method. Crucially, the result demonstrates that, when the distance between the hinges is $d = 50$~nm, the Majorana Kramers pair is stabilized in a wide region of the parameter space. 
In addition, we checked the phase diagrams for $d = 100~$nm and $200~$nm (not shown), and found no qualitative differences among these values of $d$. Quantitatively,  the green region where the Majorana Kramers pair is present becomes smaller for a larger separation $d$, as expected.

As a side remark, in the small $K_1(0)$ (strongly interacting) regime, the gaps do not flow to exact zero. This is due to the source terms 
contained in Eqs.~\eqref{Eq:RG_SourceTerm}. 
As mentioned above, in contrast to Eqs.~\eqref{Eq:RG_EffH}, these source terms contribute to the flow equations for $\tilde{\Delta}_{n}$ and $\tilde{\Delta}_{\rm c}$ in Eqs.~\eqref{Eq:RG_SourceTerm}. Therefore, in the presence of a strong interaction, the RG flows of these parameters are stopped by the hinge length of $O(\mu$m) before going to zero, leading to tiny but finite gaps. 
In plotting Fig.~\ref{Fig:PhaseDiagram_ST}, we therefore label the region in which both the renormalized gaps $\tilde{\Delta}_{\rm c}$ and $\tilde{\Delta}_{1}$ are less than 0.1 as the ``small pairing gaps'' region (marked in red color). This region then corresponds to the normal phase (no pairing) in Fig.~\ref{Fig:PhaseDiagram}. 
In conclusion, the source-term approach confirms the results presented in Fig.~\ref{Fig:PhaseDiagram} in the main text.

\subsection{VI. Feasibility of the proposed scheme}
In this Section we examine the feasibility of our proposal. Taking the first experimentally confirmed material of higher-order topological insulators--bismuth (111) nanostructures~\citesupp{Schindler:2018_S}--as an example, we estimate the initial values of the RG parameters for the effective Hamiltonian method presented in the main text, including the interaction parameter $K_1$ and the bare gap ratio $\Delta_{1}(0)/\Delta_{\rm c}(0)$. Then, we discuss suitable materials for the superconductor used for the proximity effect.

 \begin{figure}[t]
 \includegraphics[width=\linewidth]{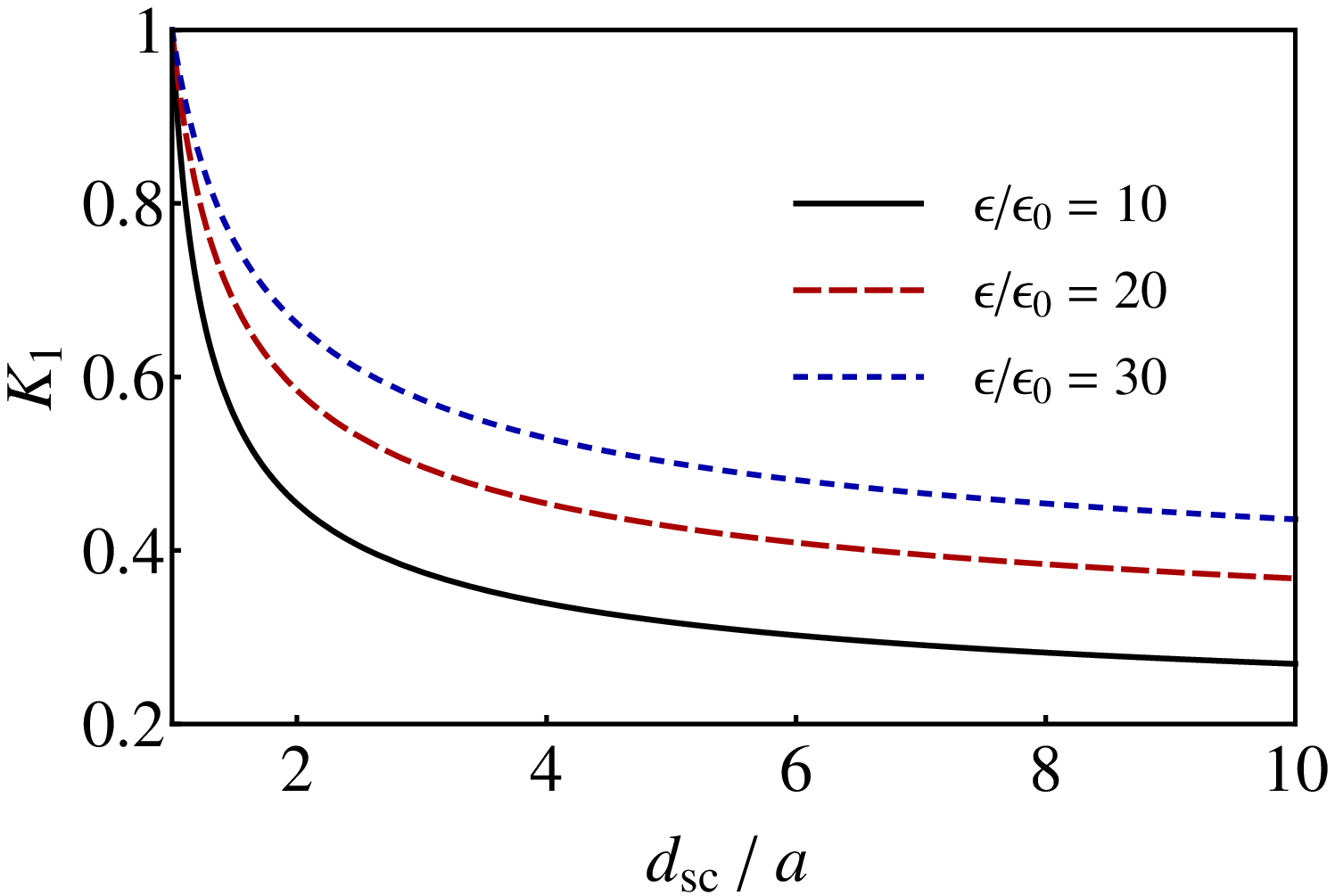}
 \caption{Interaction parameter ($K_1$) as a function of the ratio of the screening length to the hinge-state transverse decay length ($d_{\rm sc}/a$). The legend denotes various values of the relative permittivity, which is the ratio of dielectric constant ($\epsilon$) to its value in vacuum ($\epsilon_0$). The interaction parameter is computed using Eq.~\eqref{Eq:K} with the parameter $v_F = 10^5$~m/s.
}
 \label{Fig:K}
 \end{figure}

We first estimate the interaction parameter $K_1$ by modifying the formula for the helical edge states of two-dimensional topological insulators~\citesupp{Maciejko:2009_S} 
\begin{equation}
K_1 = \left[ 1 + \frac{2e^2}{\pi^2 \epsilon \hbar v_{F} } \ln \Big( \frac{d_{\rm sc}}{a}\Big)\right]^{-1/2} ,
\label{Eq:K}
\end{equation}
with the Fermi velocity $v_F$ and the transverse decay length $a$ of the hinge states. In the above, $\epsilon$ is the dielectric constant of the material, and $d_{\rm sc}$ is the screening length. We take $v_F = 10^5~$m/s using the relation $\Delta = \hbar v_F/a$, where the finite-size-induced gap $\Delta \sim O(0.1~{\rm eV})$ is obtained from band-structure calculations for Bi(111) bilayer films~\citesupp{Koroteev:2008_S,Wada:2011_S}, and the transverse decay length $a \sim 1~$nm is extracted from the current-phase relation in asymmetric SQUID experiment~\citesupp{Schindler:2018_S}.
Even though Bi bulk is a semimetal, we are interested in a Bi nanowire gapped by the finite-size effect. To estimate its dielectric constant, we take several values typical for semiconductors. The screening length $d_{\rm sc}$ depends on the experimental conditions, so we take it as a variable in the range $d_{\rm sc}/a \in [1,10]$ (we note that a larger $d_{\rm sc}$ will make the interaction parameter even smaller). 

The estimated value of the interaction parameter is plotted in Fig.~\ref{Fig:K}, which shows that the estimated $K_1$ value is typically below 0.6. We remark that the extracted value of $a \sim 1~$nm~\citesupp{Schindler:2018_S} indicates quite narrow one-dimensional conducting channels along the hinges. Concluding, the estimated value of the interaction parameter $K_1$ indicates rather strong electron-electron interactions in the hinge states due to the strong spatial confinements.

 \begin{figure}[t]
 \includegraphics[width=\linewidth]{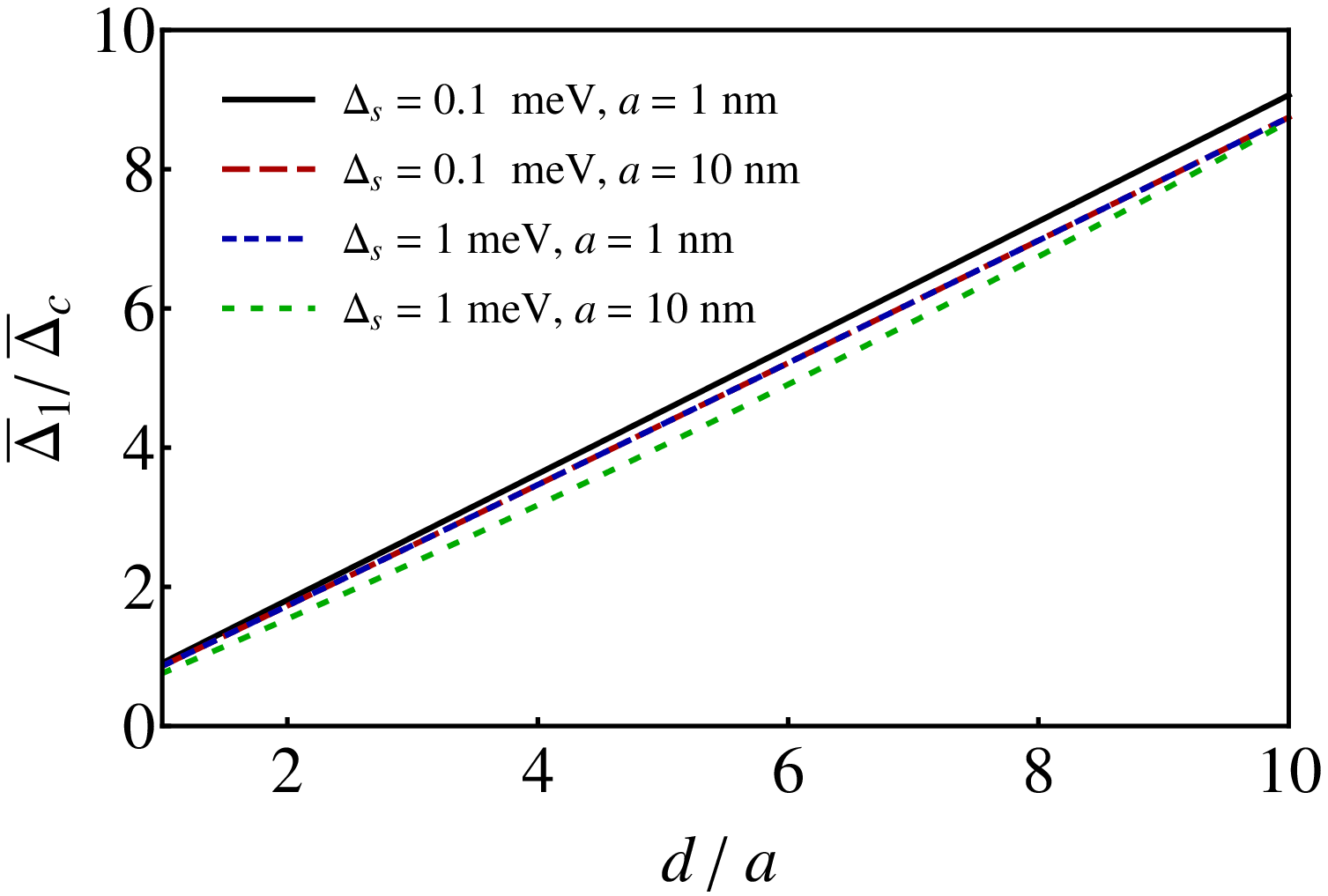}
 \caption{Bare gap ratio ($\overline{\Delta}_{1}/\overline{\Delta}_{\rm c}$) as a function of the ratio of the inter-hinge separation to the decay length ($d/a$). The legend lists various values of the superconducting gap ($\Delta_{\rm s}$) and the hinge-state transverse decay length ($a$). 
The bare gap ratio is estimated using  Eq.~\eqref{Eq:GapRatio} with $\xi_s = \hbar v_{Fs}/\Delta_{\rm s}$. 
The adopted values of the other parameters are $v_F = 10^5$~m/s and $v_{Fs}=10^6~$m/s, which gives the value $\xi_s=1.9~\mu$m for $\Delta_{\rm s}=0.35~$meV.
}
 \label{Fig:GapRatio}
 \end{figure}

Next, we estimate the bare gap ratio using the initial values of the source terms in Eqs.~\eqref{Eq:SourceTerm}, which generate the pairing gaps under the RG flow and therefore give the initial values of the gaps. Upon approximating the fast oscillating factor $|\sin (k_{Fs} d)| \approx 1$ and the velocities $u_1 \approx u_2 \approx v_F$, we get 
\begin{equation}
\frac{ \overline{\Delta}_{1}}{\overline{\Delta}_{\rm c}} \approx \frac{d}{a} \frac{ e^{d/\xi_s} K_{0} \Big( \frac{\Delta_{\rm s} a }{ \hbar v_F } \Big) }{ K_{0} \Big( \frac{\Delta_{\rm s}  a  }{2 \hbar v_F  } \Big) I_{0} \Big( \frac{\Delta_{\rm s}   a  }{2 \hbar v_F  } \Big)},
\label{Eq:GapRatio}
\end{equation}
which depends crucially on the ratio $d/a$ of the inter-hinge separation to the transverse decay length $a = a (0)$. 
To avoid possible confusion, in the left-hand side of Eq.~\eqref{Eq:GapRatio}, we use the overbar notations $\overline{\Delta}_{1}$ and $\overline{\Delta}_{\rm c}$ to refer to the initial values for the effective Hamiltonian method presented in Sec.~IV. In the right-hand side, on the other hand, we keep the same notations used for the source-term approach in Eqs.~\eqref{Eq:SourceTerm} given in Sec.~V.
For relevant parameters, we have $d/\xi_s \ll 1$ so that the bare gap ratio grows approximately linearly with the ratio $d/a$.
In Fig.~\ref{Fig:GapRatio}, we therefore plot the bare gap ratio as a function of $d/a$ for various $\Delta_{\rm s}$ and $a$ values. 
On top of the approximately linear dependence on $d/a$, the bare gap ratio depends weakly on the pairing gap of the parent superconductor $\Delta_{\rm s}$ and the decay length $a$.
Importantly, in the relevant ranges of the parameters $\Delta_{\rm s} \in [0.1~{\rm meV},~1~{\rm meV}]$ and $a \in [1~{\rm nm},~10~{\rm nm}]$, the typical value of $\overline{\Delta}_{1}/\overline{\Delta}_{\rm c}$ (i.e. the initial values $\Delta_{1}(0)/\Delta_{\rm c}(0)$ for the effective Hamiltonian method) is in the order of unity, which is compatible with the parameters adopted in our RG analysis.

Finally, we comment on suitable materials for the parent superconductor used for the proximity effect. 
The phase diagram predicted with the source-term approach (see Fig.~\ref{Fig:PhaseDiagram_ST}) indicates that Majorana Kramers pairs can be achieved using higher-order topological insulators with the inter-hinge separation $d \sim O(10$--100~nm) in the proximity of a superconductor with the gap $\Delta_{s} \sim O(0.1~$meV) and the coherence length $\xi_{s} \sim O(1~\mu$m). These material parameters suggest that aluminum is a suitable material for the parent superconductor (due to its long coherence length).
We remark that the observed ballistic supercurrent with nearly perfect transmission indicates good contacts between the hinge states and a superconductor~\citesupp{Murani:2017_S,Schindler:2018_S}, suggesting that bismuth nanowires offer a suitable platform for proximity-induced superconductivity.

In summary, together with the phase diagram presented in the main text (see Fig.~\ref{Fig:PhaseDiagram}), we conclude that Bi nanowires in the proximity of an aluminum superconductor offer a suitable platform for stabilizing Majorana Kramers pairs. We stress that the proposed scheme can be implemented in other higher-order topological insulators hosting helical hinge states, and is not restricted to the specific material. We hope that this work can stimulate studies searching for other promising materials for realization of Majorana Kramers pairs.
Given the recent rapid progress on the hunting of topological materials~\citesupp{Vergniory:2018_S}, we are optimistic that our proposal can be realized in the near future.

\let\temp\addcontentsline 
\renewcommand{\addcontentsline}[3]{} 

\bibliographystylesupp{apsrev4-1}
\bibliographysupp{MKPsupp}

\let\addcontentsline\temp

\end{document}